\documentclass[showpacs,twocolumn,pre]{revtex4-1}
\usepackage[utf8]{inputenc}
\usepackage{graphicx}
\usepackage{color}
\usepackage{appendix}
\usepackage{amsmath}
\usepackage{amssymb}
\usepackage{subfigure}
\usepackage{epsfig}
\usepackage[normalem]{ulem}

\newcommand{\cc}{\boldsymbol{c}}
\newcommand{\bb}{\boldsymbol{b}}

\newcommand{\D}{\boldsymbol{D}}

\newcommand{\G}{\boldsymbol{G}}
\newcommand{\Gama}{\boldsymbol{\Gamma}}
\newcommand{\nd}{\boldsymbol{S}}
\newcommand{\Q}{\boldsymbol{Q}}
\newcommand{\F}{\boldsymbol{F}}
\newcommand{\M}{\boldsymbol{M}}

\newcommand{\B}{\boldsymbol{B}}

\newcommand{\pp}{\boldsymbol{p}}
\newcommand{\nn}{\boldsymbol{n}}
\newcommand{\J}{\boldsymbol{J}}

\newcommand{\vv}{\boldsymbol{v}}
\newcommand{\rr}{\boldsymbol{r}}
\newcommand{\zz}{\boldsymbol{z}}

\newcommand{\xxi}{\boldsymbol{\xi}}

\newcommand{\R}{\boldsymbol{R}}

\begin{document}

\title{Anomalous fluxes in overdamped Brownian dynamics with Lorentz force}

\author{Hidde Derk Vuijk\textit{$^{1}$}, Joseph Michael Brader\textit{$^{2}$}, and Abhinav Sharma\textit{$^{1}$}}

\affiliation{\textit{$^{1}$~Leibniz Institute for Polymer Research, Hohe Str. 6, 01069 Dresden, Germany } \\
\textit{$^{2}$~University of Fribourg, Chemin du Mus\'ee 3, CH-1700 Fribourg, Switzerland}}


\begin{abstract}
We study the stochastic motion of a particle subject to spatially varying Lorentz force in the small-mass limit. The limiting procedure yields an additional drift term in the overdamped equation that cannot be obtained by simply setting mass to zero in the velocity Langevin equation. We show that whereas the overdamped equation of motion accurately captures the position statistics of the particle, it leads to unphysical fluxes in the system that persist in the long time limit; an anomalous result inconsistent with thermal equilibrium. These fluxes are calculated analytically from the overdamped equation of motion and found to be in quantitative agreement with Brownian dynamics simulations. Our study suggests that the overdamped approximation, though perfectly suited for position statistics, can yield unphysical values for velocity-dependent variables such as flux and entropy production.



\end{abstract}


\maketitle

\section{Introduction}

The motion of a particle suspended in a solvent can be modeled using the Langevin equation approach \cite{langevin1908theorie}. In this approach one writes an equation of motion for a particle, in which its interaction with the other degrees of freedom of the system (solvent) is modeled in terms of a stochastic force with suitable statistical properties. For instance, in absence of hydrodynamics the dynamics of a Brownian particle of mass $m$ can be described by the Langevin equation for its position $\rr$ and velocity $\vv$:
\begin{align}
\dot \rr(t) &= \vv(t), \nonumber \\
m\dot \vv(t) &= \F(\rr(t)) -\gamma \vv(t) + \sqrt{2\gamma k_B T}\xxi(t),
\label{Langevin}
\end{align}
where $\F(\rr)$ is an external force, $\gamma$ is a friction coefficient, $k_B$ is the Boltzmann constant and $T$ is the temperature. The noise $\xxi(t)$ is Gaussian with zero mean and time correlation $\langle\xxi(t)\xxi^{T}(t')\rangle=\boldsymbol{1}\delta(t-t')$. The importance of the Langevin equation approach lies in its applicability to a wide class of nonequilibrium problems \cite{balakrishnan2008elements}. 

The velocity correlations decay on a time scale $\tau = m/\gamma$, which implies that for times $t>>\tau$, the inertia term $m\dot \vv(t)$ is negligible and can be set to zero to obtain an effective equation of motion for $\rr$ as
\begin{align}
\gamma \dot \rr(t) = \F(\rr(t))  +\sqrt{2\gamma k_B T} \xxi(t).
\label{overdamped}
\end{align}
This equation, referred to as the overdamped equation of motion, is extensively used in theoretical and computational studies of nonequilibrium problems in which the correlation time $\tau$ is much smaller than the time scale of diffusion the particle \cite{balakrishnan2008elements,gardiner2009stochastic}. This decoupling of velocity and position on time scales larger than $\tau$ makes it easier to find analytical solutions and has the advantage of significantly faster numerical computation. In fact, it has become a common practise to start with the overdamped equation of motion of the particle (Eq. \eqref{overdamped}) as the model of the nonequilibrium system under study \cite{brader2011density,cates2013when,farage2015effective,sharma2016green-kubo,stenhammar2016light,sharma2017escape,sharma2017brownian,vuijk2018pseudochemotaxis}. 

The overdamped equation of motion is generally obtained in a simple way: set the inertia term to zero in the velocity Langevin equation and rearrange to describe the dynamics of the slow position variable. However, this procedure does not always yield the correct overdamped equation of motion, for instance when the noise is position dependent. In this case one must follow a systematic limiting procedure ($m\rightarrow 0$) of Eq. \eqref{Langevin} to obtain the appropriate overdamped equation \cite{hottovy2015smoluchowski,volpe2016effective}. This procedure yields an an additional drift term often referred to as the noise-induced drift in the literature\cite{lau2007state,hottovy2015smoluchowski,volpe2016effective,vuijk2018pseudochemotaxis}. This additional drift term is absent if one simply sets $m=0$ in the velocity Langevin equation. 

Additional drift also appears in the overdamped Langevin equation when the friction coefficient is position dependent, or more precisely, when the coefficient multiplying the velocity is position dependent. \cite{hottovy2015smoluchowski,volpe2016effective}. One particularly interesting case, which is the main focus of this paper, is that of a Brownian particle subject to Lorentz force due to spatially varying magnetic field. The Lorentz force acting on a particle can be written as an antisymmetric matrix acting on $\vv$, which, when added to the friction term $-\gamma \vv$, results in an equation with position dependent coefficient in front of $\vv$.  Lorentz force is distinct from other nonconservative forces (e.g. shear) which input energy to the system. Shear forces can drive a system out of equilibrium resulting in nonequilibrium steady states. This stands in contrast to Lorentz force. Although Lorentz force generates particle 
currents, these are purely rotational and do no work on the system, which is thus not driven out of equilibrium. Being in equilibrium, such a system has (a) a stationary density profile given by the Boltzmann distribution and (b) no fluxes. 

In this paper, we show that whereas the overdamped equation of motion for a Brownian particle in a spatially varying magnetic field accurately captures the position statistics, it leads to unphysical fluxes in the system.
We first obtain the overdamped equation from the velocity Langevin equation using existing methods and show that the trajectory from the velocity Langevin equation converges on the trajectory from the overdamped equation with decreasing mass. We then show that
for a particular choice of the spatially varying magnetic field, the overdamped equation fails to satisfy the no flux condition in equilibrium. This anomalous behaviour of the overdamped equation is the main result of this paper. We calculate these unphysical fluxes analytically from the overdamped equation, perform Brownian dynamics simulations of the overdamped equation of motion, measure the fluxes and show that they agree with the analytical predictions. 

The anomalous behaviour does not invalidate the overdamped approximation. Rather, it is a manifestation of the subtle nature of the limiting procedure that yields the overdamped equation. The overdamped equation accurately captures the statistics of the position of the particle over finite time intervals \cite{hottovy2015smoluchowski}. This is seen clearly when one considers the Fokker-Planck equation for the position variable. The Fokker-Planck equation obtained from the overdamped equation is the same as that obtained from an independent alternative route. The computing of flux, however, involves taking the small-mass limit of velocity dependent terms which may result in additional terms. For instance, it has been shown that the overdamped equation does not yield the correct entropy production in the presence of a temperature gradient \cite{celani2012anomalous, marino2016entropy, birrell2018entropy}. 


The paper is organized as follows. In Sec. \ref{LE}, we briefly describe how the overdamped Langevin equation is obtained for a charged particle in spatially varying magnetic field. In Sec. \ref{constantB}, we consider the special case of uniform magnetic field and demonstrate the existence of unusual curl-like fluxes. In Sec. \ref{wrong_langevin}, we show analytically and numerically that the overdamped equation obtained in Sec. \ref{LE} leads to unphysical fluxes in the system. The Fokker-Planck equation for the position variable is derived in Sec. \ref{small_mass_limit}. Finally we present conclusions and brief discussion in Sec. \ref{conclusions}.

\section{Langevin equation}{\label{LE}
We consider a single charged Brownian particle in a magnetic field $\B(\rr)$. The state of the particle is determined by the position vector $\rr$ and velocity $\vv$. Omitting hydrodynamic interactions, the dynamics of the particle are described by the following Langevin equation:
\begin{align}
\dot \rr(t) &= \vv(t), \nonumber \\
m\dot \vv(t) &= -\gamma \vv + q \vv \times \B(\rr) + \sqrt{2\gamma k_B T}\xxi(t),
\label{LangevinB}
\end{align}
where $m$ is the mass of the particle, $q$ is the charge,  $k_B$ is the Boltzmann constant, $T$ is the temperature and $\xxi(t)$ is Gaussian white noise with zero mean and time correlation $\langle\xxi(t)\xxi^{T}(t')\rangle=\boldsymbol{1}\delta(t-t')$. Let $\nn$ be the unit vector in the direction of the magnetic field, and $B(\rr)$ be the magnitude (i.e., $\B(\rr) = B(\rr)\nn$). We define a matrix $\M$ with elements given by $M_{\alpha \beta } = -\epsilon_{\alpha \beta \nu}n_{\nu}$, where $\epsilon_{\alpha \beta \nu}$ is the totally antisymmetric Levi-Civita symbol in three dimensions and $n_{\nu}$ is $\nu$-component of $\nn$ for the Cartesian index $\nu$. The Lorentz force can be written as $qB(\rr)\M \vv$. One can rewrite the equation in terms of the position dependent matrix $\Gama(\rr) = (\gamma \boldsymbol{1} + qB(\rr)\M)$ as
\begin{align}
m\dot \vv(t) &= -\Gama(\rr) \vv + \sqrt{2\gamma k_B T}\xxi(t).
\label{LangevinB2}
\end{align}

When one is only interested in the slow degree of freedom (i.e. the position of the particle), the simulations are generally performed using an overdamped equation of motion. This equation of motion is obtained by taking the small-mass limit of Eq. \eqref{LangevinB2}. The limiting procedure is mathematically involved and is described in detail in \cite{hottovy2015smoluchowski}. The additional drift term of the overdamped equation corresponding to Eq. \eqref{LangevinB2} is
\begin{align}
\nd(\rr)= \frac{k_B T}{\gamma}\left(\G(\rr) \nabla \G(\rr) -\nabla \G_a(\rr)\right),
\label{additional_drift}
\end{align}
where $\G_a = (\G - \G^T)/2$ is the antisymmetric part of $\G$, and $\G \equiv \gamma \Gama^{-1}$ which is
\begin{align}
\G= \boldsymbol{1} - \frac{ \gamma q B(\rr)}{ \gamma^2 +q^2B^2(\rr)}\M +  \frac{q^2B^2(\rr)}{ \gamma^2 +q^2B^2(\rr)}\M^2.
\label{invGamma}
\end{align}
It is important to note that the drift term (Eq. \eqref{additional_drift}) depends on whether the overdamped equation is interpreted in It$\rm \hat o$ or Stratonovich sense \cite{lau2007state, gardiner2009stochastic}. Equation \eqref{additional_drift} gives the additional drift in the Stratonovich interpretation of the overdamped equation which is given as
\begin{align}
\dot \rr(t) = \frac{k_B T}{\gamma} \left(\G(\rr) \nabla \G(\rr) - \nabla \G_a(\rr) \right)+ \sqrt{\frac{2k_B T}{\gamma}}\G(\rr) \xxi(t).
\label{overdampedB}
\end{align}

The small-mass limit of the Langevin Eq. \eqref{LangevinB2} involves a subtle limiting procedure which may be appreciated by noting that (a) the integration of Langevin Eq. \eqref{LangevinB2} is independent of the interpretation (It$\rm \hat{o}$ or Stratonovich) whereas that of the overdamped equation \eqref{overdampedB} is not and (b) the term $\nabla \G_a(\rr)$ cannot be eliminated by choosing a different integration calculus; that is, this term is independent of the sense in which the overdamped equation is interpreted.

The following conventions are followed throughout the article: $\partial_{\alpha}$ stands for $\partial/\partial r_{\alpha}$, where $r_{\alpha}$ is the $\alpha$-component of $\rr$. The $\alpha$-component of $\nabla \G$ is given as $(\nabla \G)_{\alpha} = \partial_{\beta} G_{\beta \alpha}$, where repeated index is summed over. Similarly $(\G \nabla \G)_{\alpha} = G_{\alpha \nu} \partial_{\beta} G_{\beta \nu}$.

\begin{figure*}[t]
 \centering
 \includegraphics[width = \textwidth]{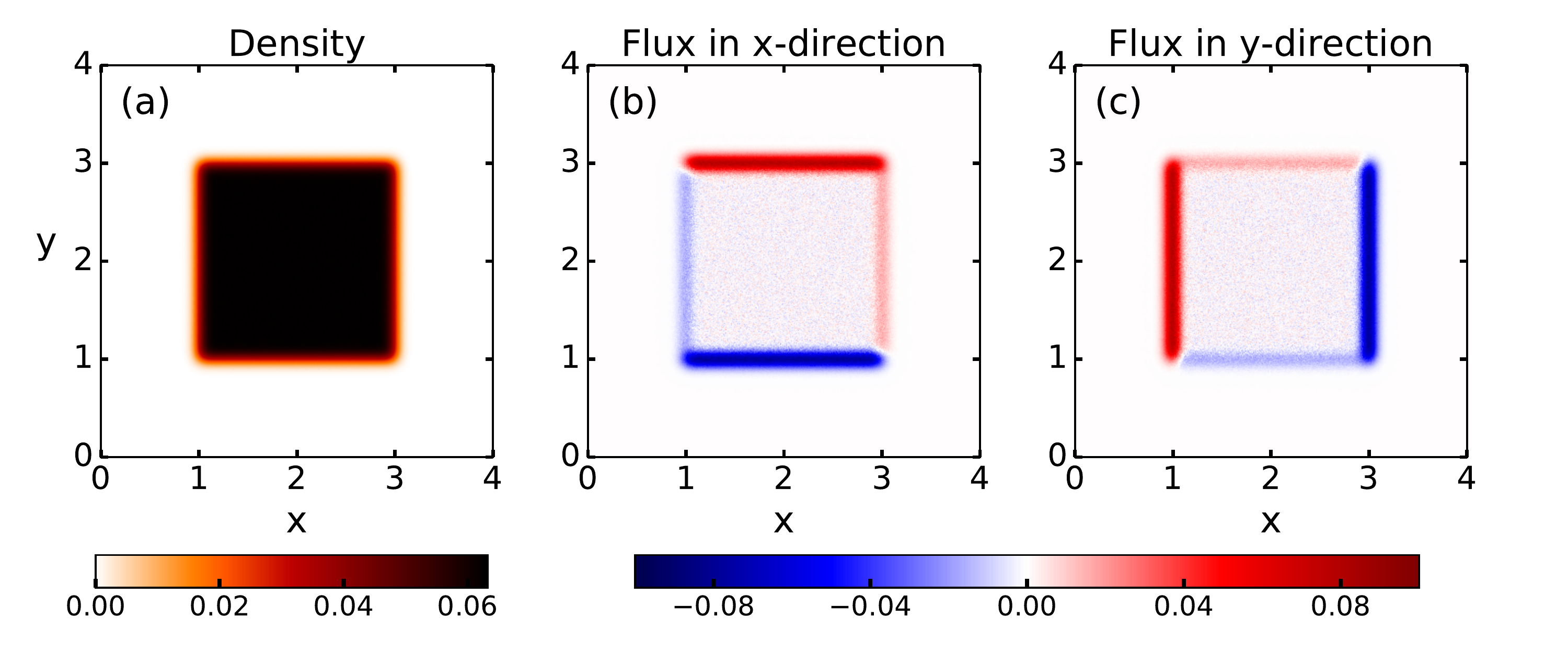}
 \caption{(a) Density distribution at time $t=0.05$ calculated by numerically integrating Eq. \eqref{LangevinB} with $m = 2\times 10^{-3}$ and $\B = 5\hat \zz$. At time $t=0$, particles are uniformly distributed in the region with $1\leq x,y \leq 3$. At $t=0.05$ density gradients exist only near the edges of the square region. (b) Flux in the $x$-direction. (c) Flux in the $y$-direction. Fluxes exist where the density gradient is large. The diffusive flux is parallel to the density gradient; the curl flux is perpendicular to the density gradient. The curl flux is along the edges of the region. This flux is divergence free and does not contribute to the time evolution of the density.}
 \label{curlflux}
\end{figure*}

\section{Uniform magnetic field}\label{constantB}
We first consider the case of uniform magnetic field. The overdamped Langevin equation can be obtained from Eq. \eqref{overdampedB} by setting $\nabla \G = 0$ as
\begin{equation}
\dot \rr(t) = \sqrt{\frac{2k_B T}{\gamma}}\G \xxi(t).
\label{uniformB}
\end{equation}
In a recent study \cite{chun2018emergence}, Chun \emph{et.} al studied the small-mass limit of the Langevin equation \eqref{LangevinB} and showed explicitly that the resulting overdamped equation is not given by Eq. \eqref{uniformB}. They first calculated the noise correlation matrix from the Langevin equation \eqref{LangevinB} for a finite mass and then took the small-mass limit to obtain the correlation matrix of the noise in the overdamped equation. This procedure yielded the surprising result that the noise appearing in the overdamped equation of motion is a nonwhite Gaussian noise. This is in sharp contrast with the overdamped equation \eqref{uniformB} which has a white Gaussian noise.

It was also shown in Ref. \cite{chun2018emergence} that the flux $\J(\rr,t)$, obtained from the correct overdamped equation, is
\begin{align}
\J(\rr,t) = -\frac{k_B T}{\gamma} \G \nabla Q(\rr,t),
\label{uniformBflux}
\end{align}
where $Q(\rr,t)$ is the probability density. This flux is unusual because the matrix $\G$ cannot be interpreted as the diffusion matrix: it is not symmetric whereas a diffusion matrix is always symmetric. The flux can be written as sum of two terms: $\G_s \nabla Q(t)$, which we call the diffusive flux determined by the symmetric part $\G_s = (\G + \G^{T})/2$ of $\G$ and $\G_a \nabla Q(t)$, which we refer to as the curl flux determined by the antisymmetric part of $\G$. We note that Eq. \eqref{uniformB} cannot give rise to curl flux. From Eq. \eqref{uniformB}, one only obtains the diffusive flux. 

The unusual flux is a consequence of the nonwhite noise that appears in the overdamped equation \cite{chun2018emergence}. Chun \emph{et.} al  studied how the nonwhite noise impacts dissipation in a system subject to nonconservative force that couples to the position. However, a direct demonstration of these fluxes was not presented. We show below that the curl flux can be measured in numerical simulations by starting with a nonequilibrium density distribution and measuring the fluxes, which arise from the density gradient. Ideally, this would be done using the overdamped equation with the nonwhite noise reported in Ref. \cite{chun2018emergence}; however, at present it is not known how to generate the nonwhite noise that appears in this equation. Therefore, we demonstrate the presence of curl flux by numerically integrating the Langevin equation \eqref{LangevinB} with a small mass. We consider noninteracting particles that are initially uniformly distributed in the region $1 \leq x,y \leq 3$ with $z=0$. We then numerically integrate the Langevin equation \eqref{LangevinB} with mass $m = 2\times 10^{-3}$,  $\B = 5\hat \zz$, and integration step $dt = 5\times10^{-6}$. Throughout the article we have used $k_B T = 1$, $\gamma = 1$, and $q=1$. The density distribution and flux are shown in Fig. \ref{curlflux} at time $t=0.05$. The velocity autocorrelation time, $m/\gamma = 2\times10^{-3}$, is much shorter than this time. As can be seen in Fig. \ref{curlflux} (a), density distribution becomes nonzero in the neighbourhood of the square region. The change in the distribution is due to the diffusive flux of the particles which is perpendicular to the edges of the square region. In addition to the diffusive flux there is also curl flux, which is shown in Figs. \ref{curlflux} (b) and (c). This flux, which is along the edges of the square region, is divergence free and therefore does not influence the time evolution of the density. 

That the flux has a curl like component has also been reported in Refs. \cite{kwon2005structure,wang2016curl}. However, the flux was obtained following the Fokker-Planck approach (shown below in Sec. \ref{small_mass_limit}) which does not require the overdamped Langevin equation. 


\begin{figure*}[t]
 \centering
 \includegraphics[width = \textwidth]{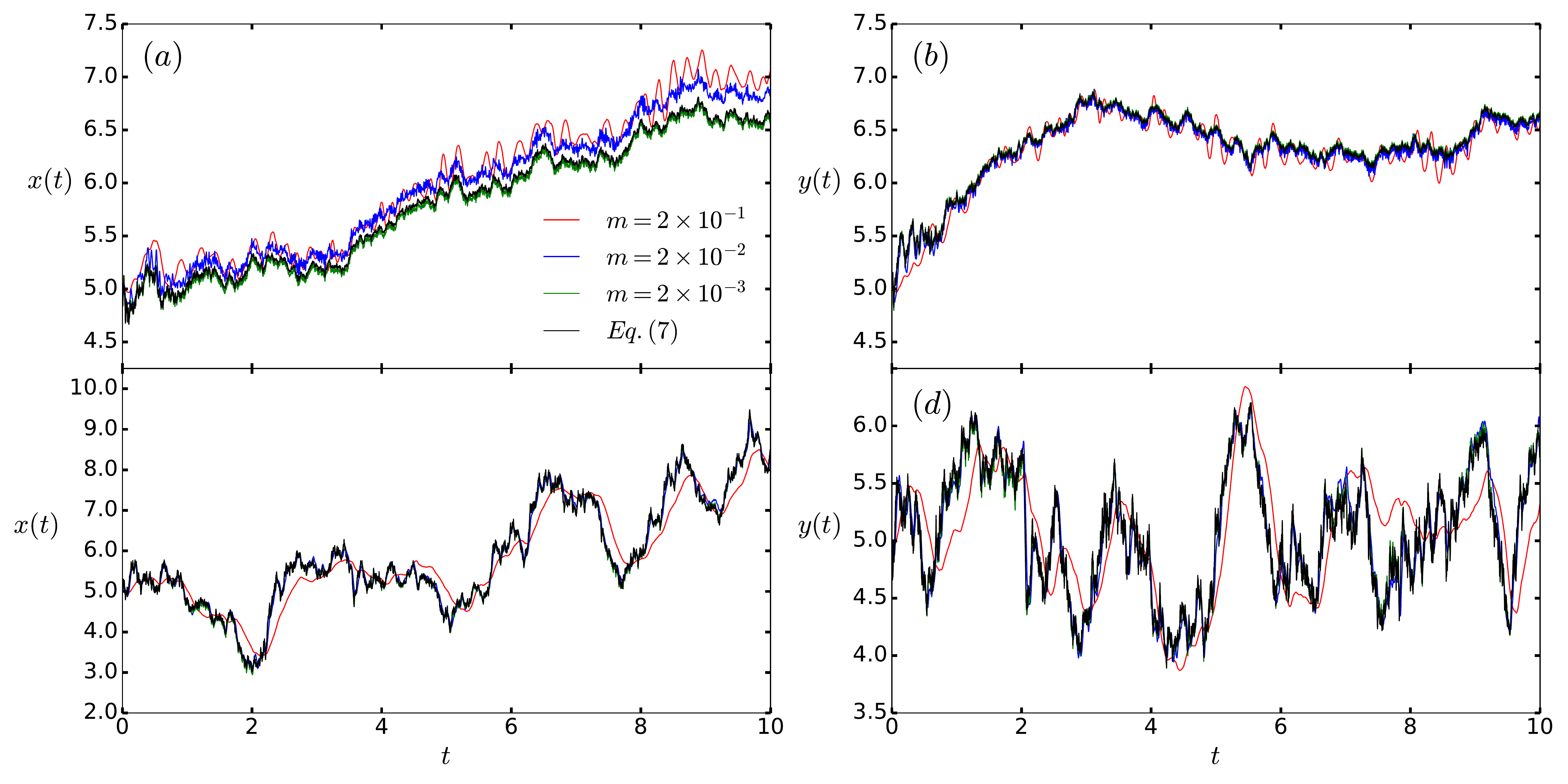}
 \caption{The $x$ and $y$ coordinates as a function of time calculated from Eq. \eqref{LangevinB} for different masses (red, blue, and green curves) and from the overdamped equation \eqref{overdampedB} (black curve). (a, b) The particle starts at $x=y=z=5$ and the magnetic field is $\B = 5 \sin(\pi y/5) \hat \zz$ . The trajectories calculated from Eq. \eqref{LangevinB} converge with the trajectory of the overdamped equation as the mass is decreased. (c, d) The particle starts at $x=y=z=5$ and the magnetic field is $1.6(y-5) $. In addition, there is a harmonic potential $(y-5)^2$. The $x$-trajectories from Eq. \eqref{LangevinB} converge with the trajectory of the overdamped equation \eqref{overdampedB} with decreasing mass. Since the magnetic field points in the $z$-direction, only the dynamics in the $x$-$y$ plane are affected by the Lorentz force.  
 }
 \label{trajectory_comparison}
\end{figure*}

\section{Inhomogeneous magnetic field}\label{wrong_langevin}
The overdamped motion of a charged particle in an inhomogeneous magnetic field has been studied in the past \cite{cerrai2011small,freidlin2012perturbations,hottovy2015smoluchowski}. In this case, the overdamped Langevin equation \eqref{overdampedB} of a particle has an additional drift term. We show below that whereas this equation accurately describes the position of the particle and therefore the correct density distribution in the long-time limit, there are fluxes in steady state.
We take the following approach: we compare the trajectory of the particle obtained from Eq. \eqref{LangevinB} with a small mass to the trajectory obtained from integrating Eq. \eqref{overdampedB}. With decreasing mass the trajectories should converge. Figure \ref{trajectory_comparison}(a,b) shows trajectories obtained from Eq. \eqref{LangevinB} with different masses and a trajectory obtained from Eq. \eqref{overdampedB}, with magnetic field $\B = 8 \sin(2\pi y/L) \hat \zz$, where $L$ is the size of the simulation box. We apply periodic boundary conditions in all directions. Figure \ref{trajectory_comparison}(c,d) shows the comparison of trajectories in presence of a harmonic potential $(y-5)^2$. The magnetic field is $\B = 1.6(y-5) \hat \zz$.  As can be seen in the Fig. \ref{trajectory_comparison}, the trajectory from  Eq. \eqref{LangevinB} seems to converge on the trajectory from Eq. \eqref{overdampedB} with decreasing mass.

Past studies have also relied on the comparison of trajectories to establish the accuracy of the overdamped Langevin equation of motion \cite{hottovy2012noise,wang2016curl,volpe2016effective}. This would seem to be a perfectly reasonable approach to establish the validity of the overdamped equation. If the two trajectories are matching, the overdamped equation of motion is accurately capturing the dynamics of the position of the particle. However, despite the matching trajectories, the two equations yield different particle fluxes in steady state. The flux obtained from the Langevin equation \eqref{LangevinB} with a small mass is identically zero at every spatial location; however, the flux obtained from the overdamped Langevin equation is nonzero in steady state; see Fig. \ref{flux_comparison}. From an equilibrium thermodynamics standpoint, the steady state should be characterized by a Boltzmann probability density with no net fluxes. The Langevin equation \eqref{LangevinB} is consistent with thermodynamic equilibrium whereas the overdamped equation \eqref{overdampedB} is not. Nevertheless, as we discuss below, this inconsistency with equilibrium does not invalidate the overdamped stochastic differential equation.

The steady state flux can be obtained analytically by evaluating
\begin{align}
\bar \J(\rr,t) = \left \langle \dot \rr (t) \delta^{(3)}(\rr (t) - \rr)\right \rangle,
\label{flux}
\end{align}
where $\rr (t)$ is the position of the particle at time $t$ and $\dot \rr (t)$ is the velocity, which is given by Eq. \eqref{overdampedB}. In order to avoid any confusion, we clarify that $\rr (t)$ is denoting the  position of the particle and $\rr$ is the position in space at which the fux is calculated. The flux can be calculated by substituting Eq. \eqref{overdampedB} for $\dot\rr (t)$ in Eq. \eqref{flux}. The term containing $\langle \xxi(t) \delta^{(3)}(\rr (t)- \rr) \rangle$ can be evaluated using the Novikov identity \cite{novikov1965ea}
\begin{align}
\langle\xi_{\alpha}(t) \R[\xxi]\rangle = \int ds \langle \xi_{\alpha}(t) \xi_{\beta}(s)  \rangle \left \langle \frac{\delta \R[\xxi]}{\delta\xi_{\beta}(s)}\right \rangle,
\end{align}
where $\xxi$ is Gaussian noise, $\alpha, \beta$ denote the $x,y$ or $z$ component, and $\R[\xxi]$ is a functional of the noise.
The details of the calculation are shown in the appendix A. The final expression for the flux is
\begin{align}
\bar{\J}(\rr,t) =-\frac{k_BT}{\gamma}\left(\nabla \G_a(\rr)\bar Q(\rr,t) + \G_s(\rr)  \nabla \bar Q(\rr,t)\right),
\label{fluxB}
\end{align}
where $\bar Q(\rr,t)$ is the probability density of the particle corresponding to the equation \eqref{overdampedB}. The second term in the expression for the flux is a diffusive flux with the position-dependent diffusion coefficient $k_B T \G_s$, where we have used $\G_s = \G \G^T$. We consider the long time limit in which the probability density is homogeneously distributed implying that the diffusive flux is identically zero. It follows from Eq. \eqref{fluxB} that for the particular choice of the magnetic field there should be a flux in the $x$-direction and no fluxes in the other directions. The $x$-component of the flux, obtained from Eq. \eqref{fluxB}, is 
\begin{align}
\bar J_x(y) = \rho_b k_B T \frac{\partial B}{\partial y}\frac{\gamma^2 - (qB)^2}{(\gamma^2 + (qB)^2)^2},
\label{jx}
\end{align}
where $\rho_b$ is the bulk probability density. Clearly, the numerically obtained flux is in excellent agreement with the analytical prediction (see Fig. \ref{flux_comparison}). 

That there are fluxes in the system is clearly inconsistent with thermal equilibrium. However, it does not invalidate the overdamped approximation. Rather, it is a manifestation of the subtle nature of the limiting procedure that yields the overdamped equation. When dealing with singular limits of equations, one can not speak broadly about the correct limiting equation in an absolute sense because the correct limit depends on the observable that one wishes to study with the limiting equation. In the case of the overdamped equation \eqref{overdampedB}, the observable is the position of the particle \cite{hottovy2015smoluchowski}. It is clear from Fig. \ref{trajectory_comparison} that indeed the overdamped equation accurately captures the position statistics of the particle. The computing of flux, however, involves taking the small mass-limit of velocity dependent terms which may result in additional contributions in the limiting procedure which are not present in the overdamped equation. Another observable which involves taking the small-mass limit of velocity dependent terms is entropy production. Previous studies have shown that the overdamped equation does not yield the correct entropy production in the presence of a temperature gradient \cite{celani2012anomalous, marino2016entropy, birrell2018entropy}.

Since the uniform magnetic field in Sec. \ref{constantB} is a special limit of the inhomogeneous magnetic field, we believe that nonwhite noise would also emerge in the case of spatially varying magnetic field. It would be ideal to obtain the correlation matrix of the noise for a spatially varying magnetic field following the same approach as in \cite{chun2018emergence}. However, at present our efforts have not been successful.

\begin{figure}[t]
 \centering
 \includegraphics[width = \columnwidth]{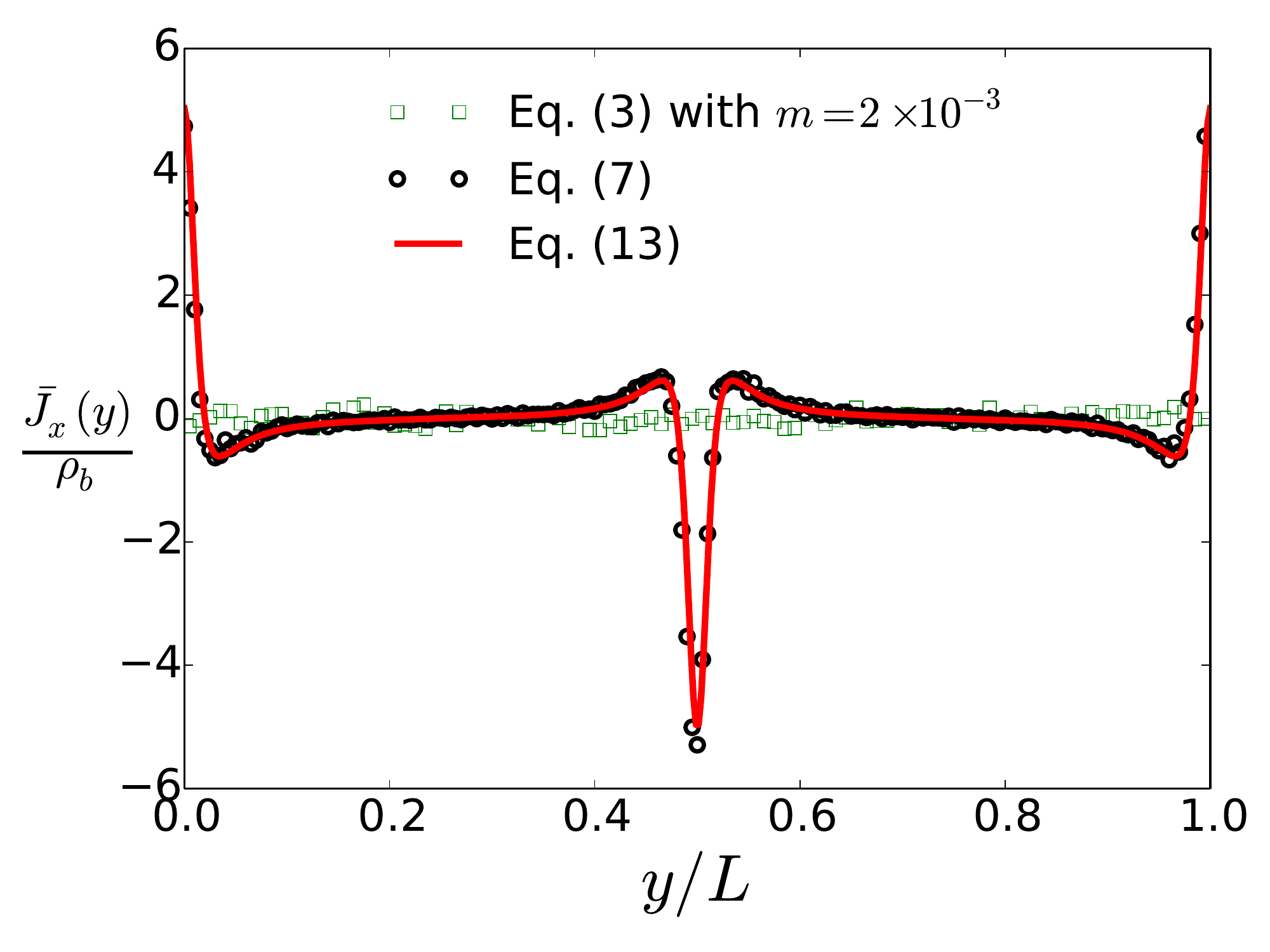}
 \caption{Steady state flux in the $x$-direction as a function of $y$ calculated from Eq. \eqref{LangevinB} (open squares) with $m = 2\times10^{-3}$, from the overdamped equation \eqref{overdampedB} (open circles), and from Eq. \eqref{jx} (red curve). The magnetic field is $\B = 8 \sin(\pi y/5) \hat \zz$. The flux is calculated by averaging the number of crossings a particle makes per unit area per unit time. The flux from the Langevin equation \eqref{LangevinB} is identically zero and the probability density is uniform. The overdamped equation violates the equilibrium condition of zero flux. The $x$ and $y$ components of flux are identically zero (not shown). }
 \label{flux_comparison}
\end{figure}

\section{Fokker Planck Equation}\label{small_mass_limit}

The Fokker-Planck equation corresponding to the overdamped equation \eqref{overdampedB} can be derived using standard methods \cite{gardiner2009stochastic} and is given as
\begin{align}
\frac{\partial}{\partial t} Q(\rr,t) = \frac{k_B T}{\gamma}\nabla \cdot \left( \G(\rr)  \nabla Q(\rr,t)\right),
\label{OFP}
\end{align}
where $\Q(\rr,t)$ is the probability density of finding the particle at location $\rr$ at time $t$. Alternatively, the Fokker-Planck equation can be obtained by an independent route which does not require the knowledge of the overdamped equation. The derivation, using the method described in Ref. \cite{risken1989fokker_ch10}, is presented in the appendix B and yields the same equation as in Eq. \eqref{OFP}. 

That one obtains the same Fokker-Planck equation establishes the validity of the overdamped equation \eqref{overdampedB} as the correct description of the position of the particle. The advantage of this alternative method of deriving the Fokker-Planck equation is that it also yields an expression for the flux $\J(\rr,t)$ in the small-mass limit:
\begin{align}
\J(\rr,t) &= -\frac{k_B T}{\gamma} \G(\rr) \nabla Q(\rr,t).
\label{JFP}
\end{align}
Though the two fluxes, Eqs. \eqref{JFP} and \eqref{fluxB} have different dependence on $\G$ and $\Q$, their divergence is the same which is why both yield the same Fokker-Planck equation. 
The steady state density distribution can be obtained from Eq. \eqref{OFP} as a uniform distribution. Consistent with thermal equilibrium, the flux in Eq. \eqref{JFP} is identically zero for a uniformly distributed density. This is in contrast to the predictions of Eq. \eqref{fluxB} which yields finite fluxes in a uniformly distributed system.


Note that the flux has exactly the same form as in Eq. \eqref{uniformBflux} but with position dependent $\G$. It may seem that one can read off the expression for flux from Eq. \eqref{OFP} by casting the Fokker-Planck equation in the form of a continuity equation $\partial Q /\partial t + \nabla \cdot \J = 0$.
Though this approach yields the correct flux in most of the cases, there can be exceptions where it would not work. For instance if the flux has a constant divergence-free part, which would leave the Fokker-Planck equation unchanged, one cannot uniquely determine the flux from the Fokker-Planck equation alone. This is clearly seen in the case of uniform magnetic field: the Fokker-Planck equation, which is given as $\partial Q /\partial t = k_B T/\gamma \nabla \cdot (\G_s \nabla Q)$,  remains unchanged due to the divergence-free flux $\G_a \nabla Q(\rr,t)$ (see Fig. \ref{curlflux}). 

%
%

\section{Discussion and conclusions}{\label{conclusions}}
In this paper we studied the motion of a Brownian particle subject to Lorentz force in the small-mass limit. We specifically considered the case in which the Lorentz force is position dependent; that is, the applied magnetic field is spatially varying. Spatially varying Lorentz force manifests itself as a position dependent coefficient in the Langevin equation for the velocity variable. One cannot then simply set the mass of the particle to zero to obtain the overdamped equation of motion \cite{hottovy2015smoluchowski}.  When the coefficient multiplying the velocity is position dependent, the small-mass limit of the Langevin equation yields an overdamped equation of motion which has an additional drift term that depends on the gradient of the coefficient. Using existing techniques, we obtained the overdamped Langevin equation of motion of the particle with the additional drift term. We compared the trajectory obtained from the overdamped equation of motion with the trajectory from the velocity Langevin equation in the limit of small mass. We found that whereas the overdamped equation of motion accurately captures the position statistics of the particle, it leads to unphysical fluxes in the system.

That there are unphysical fluxes in the system is clearly inconsistent with thermal equilibrium. However, it does not invalidate the overdamped equation \eqref{overdampedB}. The subtle limiting procedure used to obtain the overdamped equation ensures that the statistics of the position observable are accurately captured  \cite{hottovy2015smoluchowski}. However, it is not suitable for studying velocity-dependent observables such as flux. Previously, It has been shown that the overdamped equation does not yield the correct entropy production in the presence of a temperature gradient \cite{celani2012anomalous, marino2016entropy, birrell2018entropy}



The Fokker-Planck equation for the position variable can be obtained by an independent route which does not require the overdamped equation. We find that the Fokker-Planck equation obtained from this method is the same as that from the overdamped equation of motion. This establishes the validity of the overdamped equation as the corect description of the position of the particle. The flux entering the Fokker-Planck equation (Eq. \eqref{OFP}) is unusual in the sense that a density gradient gives rise not only to a flux parallel to it (diffusive) but also perpendicular to it (curl like). The unusual form of the flux in Eq. \eqref{JFP} was most recently reported in Ref. \cite{chun2018emergence} in which the authors obtained the overdamped Langevin equation of motion for a Brownian particle in a uniform magnetic field. The authors elegantly demonstrated that this equation has nonwhite noise whose correlation matrix has antisymmetric components. Unfortunately, it is presently not known how to generate such a noise process. We have not been successful to obtain the correlation matrix of the noise in the case of spatially varying magnetic field. 

Although the unusual form of flux has been previously reported, an unambiguous demonstration of such a flux using numerical simulations has been lacking. By numerically integrating the Langevin Eq. \eqref{LangevinB} with a small mass, we measured the flux directly and confirmed the theoretical predictions. When the magnetic field is uniform, the curl flux is divergence free and does not affect the time evolution of the probability distribution. By only retaining the diffusive flux, the resulting Fokker-Planck equation has a (diffusion) tensor which is symmetric. However, in a spatially varying magnetic field, the curl flux is not divergence free. Therefore, one has to retain the full tensor in the Fokker-Planck equation. This tensor can not be regarded as a diffusion tensor due to the antisymmetric components.



The Fokker-Planck equation often serves as the starting point for theoretical description of nonequilibrium problems such as spinodal decomposition \cite{archer2004dynamical,vuijk2018spinodal}, linear response \cite{sharma2016green-kubo,sharma2017brownian,merlitz2018linear}, and first passage time problems \cite{vuijk2019effect}. It will be very interesting to investigate how the presence of these unusual curl like fluxes affects the dynamics of these phenomena.

\section*{Appendix A: Calculation of flux}
Here we calculate the flux resulting from the overdamped equation of motion \eqref{overdampedB} using the Novikov relation \cite{novikov1965ea}. We denote the position of the particle at time $t$ as $\rr(t)$ to distinguish it from the spatial position $\rr$ at which we calculate the flux. The Stratonovich stochastic differential equation for the position $\rr (t)$ is given as (Eq. \eqref{overdampedB})
\begin{align}
\dot \rr (t) = \frac{k_B T}{\gamma} \left(\G \nabla\G - \nabla \G_a  \right)+ \sqrt{\frac{2k_B T}{\gamma}}\G(\rr (t)) \xxi(t).
\end{align}
The flux is calculated using $\bar \J (\rr,t) = \left \langle \dot \rr (t) \delta^{(3)}(\rr (t) - \rr)\right \rangle = \bar \J^{(1)}(\rr) + \bar \J^{(2)}(\rr)$, where $\bar \J^{(1)}$ is the contribution to the flux from the deterministic part of the equation for $\dot \rr (t)$ and $\bar \J^{(2)}$ from the stochastic part. $\bar \J^{(1)}$ can be calculated in a straightforward fashion as
\begin{align}
\J^{(1)}(\rr) &= \frac{k_B T}{\gamma} \left \langle\left(\G(\rr(t)) \nabla \G(\rr(t)) - \nabla \G_a(\rr(t))\right)\delta^{(3)}(\rr (t) - \rr) \right \rangle \nonumber \\
&=  \frac{k_B T}{\gamma}\left(\G(\rr) \nabla \G(\rr) - \nabla \G_a(\rr) \right) \bar Q(\rr,t),
\label{j1}
\end{align}
where $\bar Q(\rr,t) = \left \langle \delta^{(3)}(\rr (t) - \rr)\right \rangle$ is the probability density at $\rr$.

The calculation of $\bar \J^{(2)}$ uses the Novikov relation and is presented below. In the derivation below we have used the following \cite{fox1986functional}:
\begin{align}
 \frac{\delta}{\delta \xi_{\alpha}(t)}\int_0^t ds G_{\nu \beta}(\rr(s)) \xi_{\beta}(s) = \frac{1}{2}G_{\nu \alpha}(\rr(t)).
 \label{sweight}
\end{align}
\begin{widetext}
We calculate the flux component wise. The $\alpha$-component of the flux $\bar \J^{(2)}$ can be written as 
\begin{align}
\bar J^{(2)}_{\alpha}(\rr,t) &= \sqrt{\frac{2k_B T}{\gamma}} \left\langle G_{\alpha \beta}(\rr (t)) \xi_{\beta}(t)\delta^{(3)}(\rr (t) - \rr) \right \rangle \nonumber \\
&= \sqrt{\frac{2k_B T}{\gamma}}  \left\langle \frac{\delta}{\delta\xi_{\beta}(t)} \left(G_{\alpha \beta}(\rr (t)) \delta^{(3)}(\rr (t) - \rr) \right)\right \rangle \nonumber \\
&= \sqrt{\frac{2k_B T}{\gamma}}  \left\langle \frac{\delta r_{\nu} (t)}{\delta\xi_{\beta}(t)} \frac{\partial}{\partial r_{\nu} (t)}\left(G_{\alpha \beta}(\rr (t)) \delta^{(3)}(\rr (t) - \rr)\right) \right \rangle \nonumber \\
&= \frac{k_B T}{\gamma}\left \langle G_{\nu \beta}(\rr(t))\left(\delta^{(3)}(\rr (t) - \rr)\frac{\partial}{\partial r_{\nu} (t)}G_{\alpha \beta}(\rr (t)) + G_{\alpha \beta}(\rr (t))\frac{\partial}{\partial r_{\nu} (t)} \delta^{(3)}(\rr (t) - \rr) \right)\right \rangle \nonumber \\
&=\frac{k_B T}{\gamma} \left[\left( G_{\nu \beta}(\rr) \partial_{\nu} G_{\alpha \beta}(\rr)\right)\bar Q(\rr,t) - \partial_{\nu} \left(G_{\nu \beta}(\rr) G_{\alpha \beta}(\rr)\bar Q(\rr,t)\right) \right] \nonumber \\
&=-\frac{k_B T}{\gamma}\left[ (G_{\alpha \beta}(\rr)\partial_{\nu}G_{\nu \beta}(\rr))\bar Q(\rr,t) + G_{\alpha \beta}(\rr)G_{\nu \beta}(\rr) \partial_{\nu} \bar Q(\rr,t\right]
\label{j2}
\end{align}
where Eq. \eqref{sweight} is used in the fourth step of the derivation. Equation \eqref{j2} can be cast in vector notation as
\begin{align}
\bar \J^{(2)}(\rr,t) = -\frac{k_B T}{\gamma}\left[ \left(\G(\rr) \nabla \G(\rr)\right) \bar Q(\rr,t) + \G_s(\rr) \nabla \bar Q(\rr,t)\right]
\end{align}
Adding Eqs. \eqref{j1} and \eqref{j2}, we get
\begin{align}
\bar{\J}(\rr,t) =-\frac{k_BT}{\gamma}\left(\nabla \G_a(\rr)\bar Q(\rr,t) + \G_s(\rr) \nabla \bar Q(\rr,t)\right),
\end{align}
\end{widetext}

\section*{Appendix B: Fokker-Planck derivation}
It follows exactly from the Langevin equation \eqref{LangevinB} that the probability distribution $P(t)\equiv P(\rr,\vv,t)$ evolves in time according to \cite{gardiner2009stochastic}
\begin{align}
\frac{\partial}{\partial t}P(t) = (L_{rev} + L_{irr})P(t),
\label{FPB}
\end{align}
where the time-evolution operator has been split up in a reversible part
\begin{align}
L_{rev}P(t) = -\vv \cdot \nabla P(t) +\frac{q}{m}B(\rr)\nabla_{\vv} \cdot \left[ \M \vv P(t)\right]
\label{Lrev}
\end{align}
and an irreversible part
\begin{align}
L_{irr}P(t) = \frac{\gamma}{m} \nabla_{\vv} \cdot \left[ \vv P(t) +\frac{k_B T}{m} \nabla_{\vv} P(t)\right].
\label{Lirr}
\end{align}
To derive the Fokker-Planck equation for the position of the particle, we follow the method described in Ref. \cite{chun2018emergence, risken1989fokker_ch10}. We first recast Fokker-Planck equation equation \eqref{FPB}  as
\begin{equation}\label{small_mass_FPE_rvp_trans}
 \frac{\partial}{\partial t} \bar P(t)=\left( \bar L_{rev}+ \bar L_{irr} \right) 
 \bar P(t),
\end{equation}
where
\begin{equation}
 \bar{P}(t)=P(t)R(\vv)^{-1/2}
\end{equation}
and
\begin{equation}
 \bar{L} = R(\vv)^{-1/2} L R(\vv)^{1/2},
\end{equation}
where $L$ can be either of the operators in Eq. \eqref{FPB}, and 
\begin{equation}
 R(\vv)=\left( \frac{m}{2 \pi k_B T} \right)^{3/2} e^{-\frac{m}{2k_B T} \vv^2}
\end{equation}
is the solution to $L_{irr} R(\vv) =0$, normalized such that the integral over $\vv$ is one.
The transformed operators are
\begin{equation}
 \bar L_{irr} = -\frac{\gamma}{m}\bb^\dagger \cdot \bb,
\end{equation}
\begin{align}
  \bar L_{rev} =  -\sqrt{\frac{k_B T}{m}} \nabla \cdot \left( \bb^\dagger + \bb \right) + \frac{q}{m} \B(\rr) \cdot \left( \bb^\dagger \times \bb \right),
 \end{align}
 where 
 \begin{align}
 \bb &= \sqrt{\frac{k_B T}{m}} \nabla_{\vv} + \frac{1}{2}\sqrt{\frac{m}{k_B T}} \vv, \\\
 \bb^\dagger &= - \sqrt{\frac{k_B T}{m}} \nabla_{\vv} + \frac{1}{2}\sqrt{\frac{m}{k_B T}} \vv.
 \end{align}

 The eigenfunctions of the operator $b^{\dagger}_{\alpha} b_\alpha$, where $\alpha$ is either $x$,$y$ or $z$, are
 \begin{align}
 \psi_0(v_\alpha) = \left( \frac{m}{2 \pi k_B T} \right)^{1/4} e^{-\frac{m}{4k_B T} v_\alpha^2},
 \end{align}
 and
 \begin{align}
  \psi_n(v_\alpha) =  \frac{\psi_0(v_\alpha)}{\sqrt{n!2^n}} H_n\left( \sqrt{\frac{m}{2k_BT}} v_\alpha \right),
 \end{align}
where  $H_n$ are Hermite polynomials.
The operators $b_\alpha^\dagger$ and $b_\alpha$ are the raising and lowering operators of the eigenfunctions:
$b_\alpha^\dagger \psi_n(v_\alpha)=\sqrt{n+1} \psi_{n+1}(v_\alpha)$ and
$b_\alpha \psi_n(v_\alpha)=\sqrt{n} \psi_{n-1}(v_\alpha)$.
The eigenfunctions  are orthonormal,
\begin{equation}
 \int_{-\infty}^{\infty} dx \psi_n(x) \psi_m(x) = \delta_{n,m},
\end{equation}
 and can be used to expand $\bar P(t)$:
\begin{equation}\label{small_mass_P_expansion}
 \bar P(t) = \sum_{n_x,n_y,n_z = 0}^\infty c_{n_x,n_y,n_z} \psi_{n_x}(v_x) \psi_{n_y}(v_y) \psi_{n_z}(v_z),
\end{equation}
where $c_{n_x,n_y,n_z} = c_{n_x,n_y,n_z}(\rr,\pp,t)$.

Without loss of generality, the magnetic field is oriented along the $z$ direction and $\B(\rr) = B(\rr) \hat \zz$.
Equation \eqref{small_mass_FPE_rvp_trans} together with the orthonormality of the eigenfunctions yields an hierarchy of equations 
for the functions $c_{n_x,n_y,n_z}$ called a Brinkman hierarchy \cite{brinkman1956brownian}:
\begin{widetext}
 \begin{align}\label{small_mass_brinkman}
  \frac{\partial}{\partial t} c_{n_x,n_y,n_z} =& -\frac{\gamma}{m} c_{n_x,n_y,n_z}(n_x+n_y+n_z) - \D \cdot 
    \begin{bmatrix}
     \sqrt{n_x+1} c_{n_x+1,n_y,n_z}\\
     \sqrt{n_y+1} c_{n_x,n_y+1,n_z}\\
     \sqrt{n_z+1} c_{n_x,n_y,n_z+1}
    \end{bmatrix}
    - \D \cdot
    \begin{bmatrix}
     \sqrt{n_x} c_{n_x-1,n_y,n_z}\\
     \sqrt{n_y} c_{n_x,n_y-1,n_z}\\
     \sqrt{n_z} c_{n_x,n_y,n_z-1}
    \end{bmatrix} \nonumber \\
   & + \frac{qB(\rr)}{m} \sqrt{n_x(n_y+1)} c_{n_x-1,n_y+1,n_z}
    - \frac{qB(\rr)}{m} \sqrt{(n_x+1)n_y} c_{n_x+1,n_y-1,n_z},
 \end{align}
\end{widetext}
where $\D = \sqrt{\frac{k_B T}{m}} \nabla$.

The probability density for the position and orientation, $Q(t)\equiv Q(\rr,t)$, is given by the 
first expansion coefficient:
\begin{align}
 Q(t) =& \int d \vv P(t) \nonumber \\
 =& \int d \vv \bar P(t) \psi_0(v_x)\psi_0(v_y)\psi_0(v_z) \nonumber \\
 =& c_{0,0,0}.
\end{align}
The order of the coefficient functions $c_{n_x,n_y,n_z} = \mathcal{O}(m^{\frac{1}{2}(n_x+n_y+n_z)}) $
and up to leading order in $m$, $\partial_t c_{n_x,n_y,n_z} =0$ for $n_x+n_y+n_z > 0$.
Up to leading order in $m$ Eq. \eqref{small_mass_brinkman} is closed and can now be written as
\begin{align}
 \frac{\partial}{\partial t} c_{0,0,0} = - \sqrt{\frac{k_B T}{m}} \nabla \cdot \cc_1,
 \end{align}
 and
 \begin{align}
\Gamma \cc_1 = - \sqrt{m k_B T} \nabla c_{0,0,0},
\end{align}
where  
\begin{equation*}
\cc_1 = 
    \begin{bmatrix}
     c_{1,0,0} \\
     c_{0,1,0} \\
     c_{0,0,1}
    \end{bmatrix}
\text{~and~}
 \Gamma = 
     \begin{bmatrix}
	\gamma & -qB(\rr) & 0 \\
	qB(\rr) & \gamma & 0 \\
	0 & 0& \gamma
    \end{bmatrix}.
\end{equation*}
The matrix $\Gamma$ is the sum of $\gamma \boldsymbol{1}$ and the cross product with $\B(\rr)$, 
where in this case $\B(\rr)=B(\rr) \hat \zz$.
In the general case of magnetic field as $\B(\rr) = B(\rr)\nn$ where $\nn$ is a unit vector, the friction matrix is given as $\Gama(\rr) = (\gamma \boldsymbol{1} + qB(\rr)\M)$. The elements of the matrix $\M$ are given as $M_{ \alpha \beta } = -\epsilon_{\alpha \beta \nu}n_{\nu}$, where $\epsilon_{\alpha \beta \nu}$ is the totally antisymmetric Levi-Civita symbol in three dimensions and $n_{\nu}$ is $\nu$-component of $\nn$ for the Cartesian index $\nu$.

The flux in position space is
\begin{align}
 \J(\rr,t) &= \int d\vv ~ \vv P(\rr,\vv,t)
\end{align}
which can be calculated by using Eq. \eqref{small_mass_P_expansion} and 
$v_{\alpha} \psi_0(v_\alpha) = \sqrt{\frac{T}{m}} \psi_1(v_\alpha)$:
\begin{align}
 \J(\rr,t) &= \sqrt{\frac{k_B T}{m}} \cc_1(\rr,t)  \nonumber \\
 &= - k_B T \Gamma^{-1}  \nabla Q(t)
\end{align}
So the equation for the probability density $Q(t) \equiv Q(\rr,t)$ is
\begin{align}\label{small_mass_FPE_Q}
 \frac{\partial}{\partial t} Q(t) &=  -\nabla \cdot \J(\rr,t) \nonumber \\
 &= k_B T\nabla   \cdot \left(\Gamma^{-1} \nabla Q(t)\right),
\end{align}
where $\Gamma^{-1}(\rr)$ is given by Eq. \eqref{invGamma}.

\bibliographystyle{apsrev4-1}

%


\end{document}